\documentclass[%
 aip,
amsmath,amssymb,
preprint,
reprint, 
]{revtex4-2}

\usepackage{graphicx}
\usepackage{dcolumn}
\usepackage{bm}

\usepackage[utf8]{inputenc}
\usepackage[T1]{fontenc}
\usepackage{mathptmx}
\usepackage{etoolbox}
\usepackage{xcolor}

\makeatletter
\def\@email#1#2{%
 \endgroup
 \patchcmd{\titleblock@produce}
  {\frontmatter@RRAPformat}
  {\frontmatter@RRAPformat{\produce@RRAP{*#1\href{mailto:#2}{#2}}}\frontmatter@RRAPformat}
  {}{}
}%
\makeatother
\begin{document}

\preprint{AIP/123-QED}

\title{Rydberg electromagnetically induced transparency based laser lock to Zeeman sublevels with 0.6 GHz scanning range}
\author{Alexey Vylegzhanin}
\email{Alexey.Vylegzhanin@oist.jp}
 \affiliation{Okinawa Institute of Science and Technology Graduate University, Onna, Okinawa 904-0495, Japan.}

\author{Síle {Nic Chormaic}}
\email{Sile.NicChormaic@oist.jp}
\affiliation{Okinawa Institute of Science and Technology Graduate University, Onna, Okinawa 904-0495, Japan.}
\author{Dylan J. Brown}
\affiliation{Okinawa Institute of Science and Technology Graduate University, Onna, Okinawa 904-0495, Japan.}

\date{\today}

\begin{abstract}
We propose a technique for frequency locking a laser to the Zeeman sublevel transitions between the $\mathrm{5P}_{3/2}$ intermediate and $\mathrm{32D}_{5/2}$ Rydberg states in $\mathrm{^{87}Rb}$. This method allows for continuous frequency tuning over 0.6 GHz by varying an applied external magnetic field. In the presence of the applied field, the electromagnetically induced transparency (EIT) spectrum of an atomic vapor splits via the Zeeman effect according to the strength of the magnetic field and the polarization of the pump and probe lasers. We show that the 480~nm pump laser, responsible for transitions between the Zeeman sublevels of the intermediate state and the Rydberg state, can be locked to the Zeeman-split EIT peaks. The short-term frequency stability of the laser lock is 0.15 MHz and the long-term stability is within 0.5 MHz. The linewidth of the laser lock is $\sim$0.8~MHz and $\sim$1.8~MHz in the presence and in the absence of the external magnetic field, respectively. In addition, we show that in the absence of an applied magnetic field and adequate shielding, the frequency shift of the lock point has a peak-to-peak variation of 1.6~MHz depending on the polarization of the pump field, while when locked to Zeeman sublevels this variation is reduced to 0.6 MHz. The proposed technique is useful for research involving Rydberg atoms, where large continuous tuning of the laser frequency with stable locking is required.

\end{abstract}

\maketitle
\section{Introduction}
Many present-day atomic, molecular  or precision spectroscopy experiments require a frequency-stabilized laser source, engineered to some extent via the frequency and the linewidth to address a particular atomic or molecular transition\cite{bendkowsky2009observation, jiang2011making, mcgrew2018atomic}. Frequently, optical absorption~\cite{pearman2002polarization},  optical cavities\cite{drever1983laser, jiang2011making}, or frequency combs\cite{picque2019frequency} are used as a frequency reference to stabilize the laser. Some of the most common techniques for  frequency locking include saturated absorption spectroscopy (SAS)~\cite{schmidt1994cesium}, Pound-Drever-Hall (PDH) locking~\cite{black2001introduction}, modulation transfer spectroscopy (MTS)~\cite{mccarron2008modulation}, fluorescence spectroscopy~\cite{nieddu2019simple}, and dichroic atomic vapor laser locking (DAVLL)~\cite{corwin1998frequency}. Another technique, electromagnetically induced transparency (EIT) locking~\cite{boller1991observation,lukin1998intracavity,fleischhauer1999electromagnetically, rajasree20211}, can be exploited to lock the relative frequency of two or more lasers. This method is useful if one needs to access a narrow transition when a direct absorption signal is not feasible, for example, in the excitation of neutral atoms to Rydberg states~\cite{mohapatra2007coherent, naber2017electromagnetically,zhang2018interplay,jia2020frequency,rajasree2020generation, su2022optimizing, vylegzhanin2023excitation}. The particular interest in Rydberg atoms lies in their potential as a platform for quantum computation~\cite{demonstration_of_C_NOT_gate, quantum_information_Saffman, adams2019rydberg_quantum_tech, chew2022ultrafast}, quantum simulation~\cite{weimer2010rydberg_simulator, bharti2022ultrafast}, and quantum repeater~\cite{zhao2010efficient} technologies, due to their particular properties, such as the long lifetime of  Rydberg states, strong dipole-dipole interactions, and the Rydberg blockade phenomenon~\cite{jaksch2000fast, gaetan2009observation, pritchard2010cooperative,petrosyan2011electromagnetically}. In most related experiments, reliable laser locking and frequency scanning techniques are needed to  precisely control optical transitions to the Rydberg states, as well as between  Zeeman sublevels for the purpose of state preparation. 

Some experiments require trapped Rydberg atoms and most demonstrated traps rely on a magnetic potential~\cite{choi2005magnetic,anderson2013production,hermann2014long}, the shape and depth of which strongly depends on the $m_J$ quantum number, where $m_J$ represents the Zeeman sublevels, each requiring a different transition wavelength. Hence, having control over Rydberg-state Zeeman sublevels, and, therefore, any associated magnetic potential, requires a reference transition frequency for each of the sublevels. Rydberg atoms are also viewed as excellent tools for microwave sensing applications~\cite{kumar2017rydberg, robinson2021determining, menchetti2023digitally, li2023magnetic} due to their high sensitivity to electric fields. Controlling the microwave frequency detection range necessitates managing the optical transitions to  Rydberg states as well as the separation between different Rydberg states or the various Zeeman sublevels within a particular Rydberg state~\cite{shi2023tunable}.

\begin{figure*}
    \centering
    \includegraphics[width=17cm]{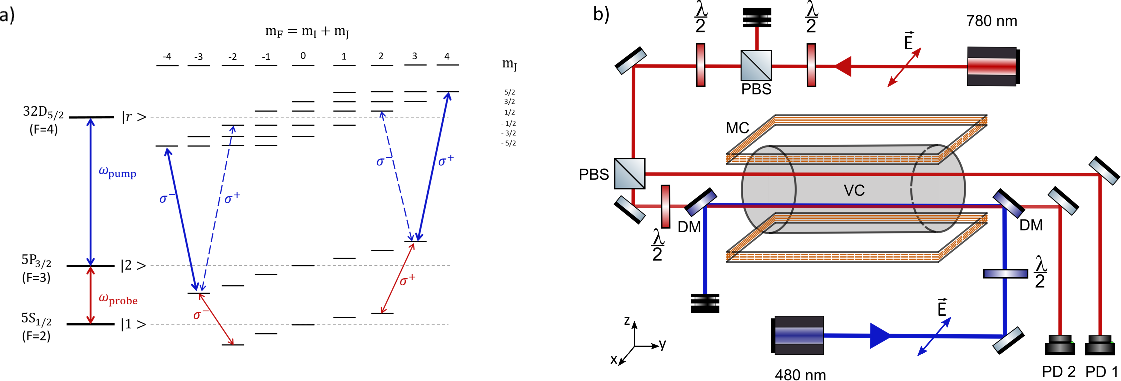}
    \caption{a) Energy level diagram for the Zeeman-split Rydberg EIT transitions. The 780~nm probe light ($\omega_{\mathrm{probe}}$) and the 480~nm pump light ($\omega_{\mathrm{pump}}$) are linearly polarized perpendicular to the magnetic field, therefore seen as a linear combination of $\sigma^+$ and $ \sigma^-$ polarized light by the atoms. b) Schematic of the experiment for excitation to Rydberg states in a vapor cell. PD - photodetector, VC - vapor cell, MC - magnetic coils, PBS- polarizing beam splitter, DM - dichroic mirror. The $\lambda/2$ plate in the path of the 780~nm beam controls the polarization of the 780~nm light for optical pumping. The $\lambda/2$ plate in the path of the 480~nm beam is used to access different $32\mathrm{D}_{5/2}$ Rydberg  $m_J$ states.}
    \label{fig:setup}
\end{figure*}

In this work, we demonstrated laser locking to Zeeman sublevel EIT peaks in a $\mathrm{^{87}Rb}$ three-level, cascaded Rydberg system, in the presence of an applied magnetic field (see Fig.~\ref{fig:setup}(a)), in a manner analogous to the work by Bao \textit{et al.}~\cite{bao2016tunable} for Cs atoms. Our method allowed us to continuously shift the frequency of the 480~nm pump laser ($\omega_{\mathrm{pump}}$) by up to by up to 300~MHz either side of the locking position with zero applied magnetic field. We also demonstrated that, when the Earth's and stray magnetic fields were not shielded from the system, the reference, single-peak, EIT signal experienced a frequency shift due to the difference in the energy shift between two neighboring $m_J$ states. The nondegenerate Zeeman sublevels were indistinguishable on the observed EIT signal. However, depending on the polarization of the 480~nm pump, the locking position shifted according to the most dominant $m_J$ transition. This introduced a frequency uncertainty of 1.6~MHz between the extreme locking points which could be detrimental for spectroscopy or precision measurement experiments.

\section{Experimental setup}
The experimental setup is shown in Fig.~\ref{fig:setup}(b). We used a room-temperature $\mathrm{^{87}Rb}$-enriched vapor cell (TT-RB87-75-V-P, TRIAD Technology Inc.) of diameter $d=25$~mm and length $l=75$~mm. An applied DC magnetic field was produced along the $z$-axis, transverse to the light beam propagation axis, by a pair of rectangular coils with dimensions $\mathrm{150~mm\times70~mm}$ and $N=50$ turns connected in series. Each coil was approximately $\mathrm{30~mm}$ from the center of the cell. The magnetic field was measured at the center of the cell by a gaussmeter (HIRST Magnetic Instruments Ltd., GM08) for currents varying from $I=$2.0 to 6.0~A with 0.1 Gauss precision. The 780~nm laser (DL pro, Toptica), used as the probe in the two-photon EIT scheme (see Fig.\ref{fig:setup}(a)) was locked via SAS to the $\mathrm{^{85}Rb}$ $ \left|\mathrm{5S}_{1/2}, F=3\right> \rightarrow \left|\mathrm{5P}_{3/2}, F'=3,4\right>$ crossover peak. It was then shifted by 1.0662~GHz using an electro-optic modulator (EOM, NIR-NPX800 LN-10, Photline Technologies) to be resonant with the $\mathrm{^{87}Rb}$ $ \left|\mathrm{5S}_{1/2}, F=2\right> \rightarrow \left|\mathrm{5P}_{3/2}, F'=3\right>$ transition. For the purposes of this work, the EOM was not used to tune the lock, and was driven with a fixed frequency of 1.0662~GHz.  The  480~nm pump  light was derived from a frequency-doubled, high power 960~nm laser (TA-SHG pro, Toptica). Approximately 1~$\mu$W of the 960~nm seed light was collected prior to the doubling cavity and was sent via an optical fiber to a wavemeter (HighFinesse Ångstrom WS-6/600) to measure the laser wavelength. Each wavelength measurement was obtained by taking an average of at least 200 values and then converted into frequency for the 480~nm light.

\begin{figure}[ht!]
    \centering
    \includegraphics[width=8.5cm]{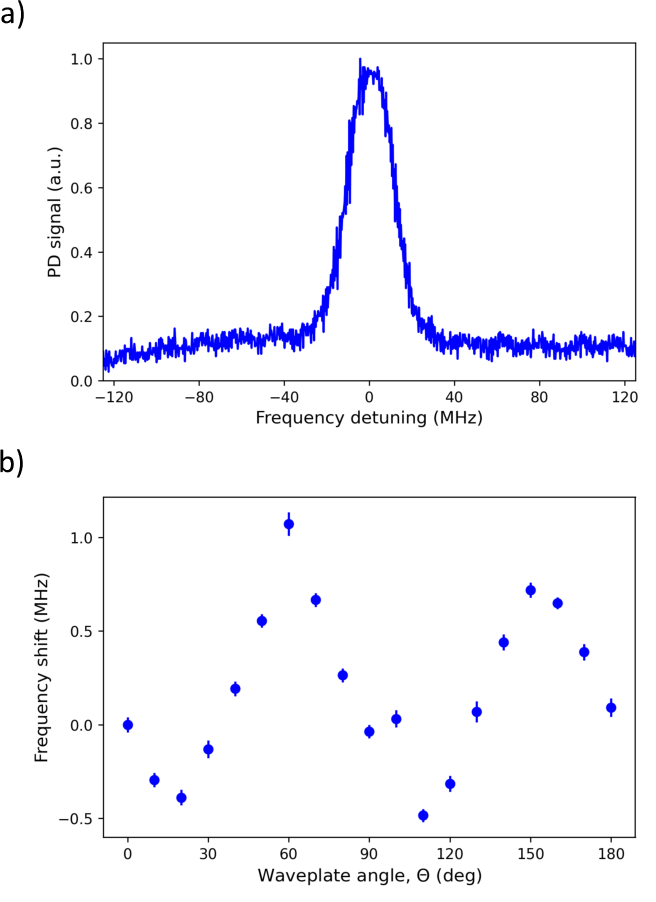}
    \caption  {a) Normalized spectrum of the EIT peak in the absence of an applied magnetic field. The different Zeeman levels are indistinguishable from one another. b) Dependence of the EIT locking frequency shift from the reference point on the polarization of the 480~nm laser light in the absence of an applied magnetic field for the $\mathrm{^{87}Rb}$ $\left| 5\mathrm{P}_{3/2} \right> \rightarrow \left| 32\mathrm{D_{5/2}} \right>$ transition. The reference point is the frequency of the 480~nm laser light when locked at $\theta=0^{\circ}$. The frequency values and the error bars were calculated from the average and the standard deviation of $\sim$200 values of the wavemeter recording, respectively. }
    \label{fig:no_mj_waveplte}
\end{figure}

The power of the 780~nm laser was set to 300~$\mu$W and that of the 480~nm laser was 300~mW. The 780~nm beam was split with a polarizing beam splitter, with one part sent through the vapor cell as a reference and the other counter-propagated with the 480~nm pump through the cell, thereby coupling the  $\left|\mathrm{5S}_{1/2}, F=2\right>$ ground, $\left|\mathrm{5P}_{3/2}, F'=3\right>$ intermediate, and $\left|\mathrm{32D}_{5/2}\right>$ Rydberg states. In the presence of the applied magnetic field, the Zeeman sublevels were nondegenerate and are shown as $m_F$ for the ground and intermediate states, and as $m_J$ for the Rydberg state (see Fig.\ref{fig:setup}(a)). A $\lambda/2$ waveplate in the beam of the 480~nm laser controlled its polarization to satisfy the momentum conservation requirement, $\Delta m_F=\pm1$, for transitions to various $m_J$ levels of the Rydberg state. When the polarization was aligned along the $x$-axis, i.e., orthogonal to the applied magnetic field, the atoms experienced a linear combination of $\sigma^+$ and $\sigma^-$ fields. When the polarization was aligned along the $z$-axis, i.e., parallel to the applied magnetic field, the atoms experienced $\pi$ polarized light. The 780~nm laser light polarization was set orthogonal to the applied magnetic field, optically pumping the atoms to the $m_F=\pm4$  Zeeman sublevels~\cite{zhang2018interplay}. The two 780~nm beams were incident on a dual photodetector (Thorlabs PDB210A/M), where the reference, incident on PD1, was subtracted from the probe, incident on PD2, to get the EIT signal. The processed signal from the photodetectors was transmitted to the Toptica Digilock 110 module and monitored using the DigiLock software. As the 480 nm laser frequency was scanned, a spectrum was obtained. When the 480~nm laser was resonant with the Rydberg transition in the absence of the applied magnetic field, a single EIT peak was visible, as shown on the measured spectrum in Fig.~\ref{fig:no_mj_waveplte}(a).

\section{Results}
\begin{figure*}
    \centering
    \includegraphics[width=17cm]{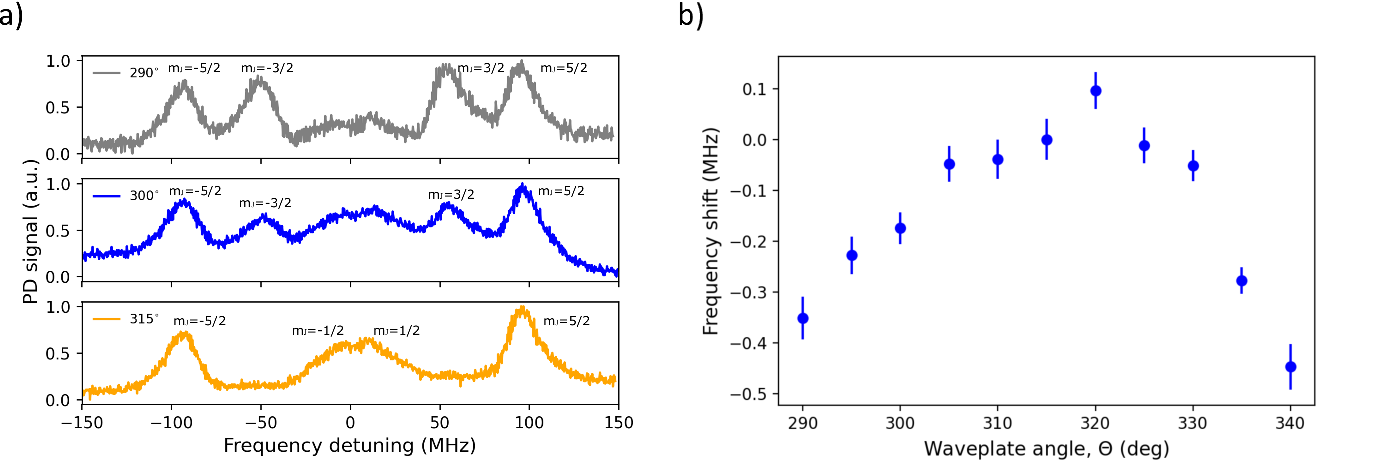}
    \caption{a) Normalized spectrum of the Zeeman EIT peaks with $\mathrm{B\approx26~G}$ and 480~nm $\lambda/2$ waveplate angles $\theta=290^{\circ}$ (gray), $\theta=300^{\circ}$ (blue) and $\theta=315^{\circ}$ (orange). b) Dependence of the EIT locking frequency shift on the polarization of the 480~nm laser light, when locked to the $ \left| 5\mathrm{P}_{3/2}, m_{F'}=3 \right> \rightarrow \left| 32\mathrm{D}_{5/2},m_J=-5/2 \right>$ transition with applied magnetic field $B=26$~G. The zero frequency point is chosen to be at waveplate angle $\theta=315^{\circ}$, since at this polarization the $m_J=-5/2$ peak has the highest amplitude. The frequency values and the error bars were calculated from the  average and standard deviation of $\sim$200 values of the wavemeter recording, respectively. }
    \label{fig:waveplate_mj_scan}
\end{figure*}

In the initial set of experiments no magnetic field was applied to the system, and the 480~nm laser was locked to the EIT signal recorded from the photodetector, as shown in Fig.~\ref{fig:no_mj_waveplte}(a). In this case, the Zeeman shifts from the Earth's magnetic field and any additional stray magnetic fields were not enough for individual Zeeman levels to be distinguishable. We then rotated the 480~nm $\lambda/2$ waveplate in steps of $10^\circ$ to change the polarization and recorded the wavelength, which we subsequently converted to  frequency values (see Fig.~\ref{fig:no_mj_waveplte}(b)).  We observed a polarization-dependent frequency shift of the lock position  with a peak-to-peak variation of 1.6~MHz. As we adjusted the 480~nm polarization, the dominant $m_J$ Rydberg level also changed according to the angular momentum conservation selection rules $\Delta_{m_F}=0,\pm1$ for $\pi$ and $\sigma^{\pm}$ polarization respectively, $\Delta_{m_F}=m_I+\Delta_{m_J}$, thereby changing the weighting of the transitions that determined the lock point. From these measurements it can be seen that, in an environment with no magnetic shielding and zero applied magnetic field, the locking frequency can vary quite significantly depending on the laser polarization.

We next applied a magnetic field of 26~G to the system using the same laser conditions as before.  We observed a splitting of the EIT peak (see Fig.~\ref{fig:waveplate_mj_scan}(a)) corresponding to the different $m_J$ levels. Depending on the polarization of the 480~nm laser, we  observed different combinations of the $m_J$ peaks with different amplitudes. We locked the 480~nm laser to the $\left| 5\mathrm{P}_{3/2}, m_{F'}=3 \right> \rightarrow \left| 32\mathrm{D}_{5/2},m_J=-5/2 \right>$ transition and varied its polarization by rotating the $\lambda/2$ waveplate, see Fig.~\ref{fig:waveplate_mj_scan}(b). One can see that the relative shift of the locking frequency is within $\sim0.6$~MHz, which is almost three times less than when no magnetic field is applied (Fig. \ref{fig:no_mj_waveplte}(b)). The variation arose from the fact that, when the polarization was changed, the amplitude of the $ \left| 5\mathrm{P}_{3/2}, m_{F'}=3 \right> \rightarrow \left| 32\mathrm{D}_{5/2},m_J=-5/2 \right>$ transition peak changed ~\cite{schlossberger2024zeeman}. At the same time, the neighboring $ \left| 5\mathrm{P}_{3/2}, m_{F'}=3 \right> \rightarrow \left| 32\mathrm{D}_{5/2},m_J=-3/2 \right>$ transition peak appeared and its tail overlapped with the locking transition peak. This affects the locking frequency. If the scan of the waveplate angle, $\theta$, were outside the range of 290$^{\circ}$ to 340$^{\circ}$ the $m_J=-5/2$ EIT peak signal disappeared and the lock was no longer possible. For the specific case where the 480~nm polarization was set such that the atoms experienced a linear combination of $\sigma^+$ and $\sigma^-$ circular polarizations, a suppression of the transitions to the $m_J=\pm3/2$ sublevels was observed (see Fig.~\ref{fig:waveplate_mj_scan}(a)) and the laser could be locked precisely to the $m_J=\pm5/2$ transition frequency.

\begin{figure*}
    \centering
    \includegraphics[width=17cm]{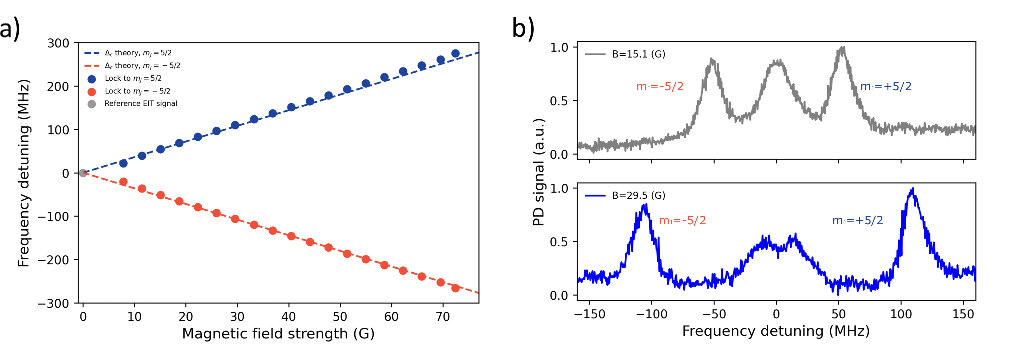}
    \caption{a) Measured frequency shift of the laser locked to $\left|  5\mathrm{P}_{3/2}, m_{F'}=3 \right> \rightarrow \left| 32\mathrm{D}_{5/2},m_J=5/2 \right>$ (dark blue) and $\left| 5\mathrm{P}_{3/2}, m_{F'}=3 \right> \rightarrow \left| 32\mathrm{D}_{5/2},m_J=-5/2 \right>$ (dark red) transitions, with the corresponding predicted detuning of the pump field $\Delta_\mathrm{c}$. Reference locking to the degenerate $\mathrm{32D_{5/2}}$ state is shown by the grey dot. b) Spectrum of the locking signal with an applied magnetic field of $\mathrm{B=15.1~G}$ (grey) and $\mathrm{B=29.5~G}$ (blue). The EIT peaks of the $m_J=\pm1/2$ Zeeman sublevels are not distinguishable within the central peak when $\mathrm{B=15.1~G}$), and are barely distinguishable by the split of the central peak when $\mathrm{B=29.5~G}$.}
    \label{fig:lock}
\end{figure*}

We  also used the applied magnetic field to control the 480~nm laser frequency similar to the work by Bao \textit{et al.} for Cs atoms~\cite{bao2016tunable}. By continuously adjusting the applied magnetic field, we could measure the change in the locking frequency of the $m_J$ peak due to the Zeeman shift. The applied magnetic field affected the ground state, the intermediate state, and the Rydberg state energy levels in the following way
\begin{equation}\label{eq:levels}
\begin{split}
    & \Delta E_r=\frac{\mu_B\cdot B}{\hbar}\cdot g_J m_J \\
    & \Delta E_2=\frac{\mu_B\cdot B}{\hbar}\cdot g_{F'} m_{F'} \\
    & \Delta E_1=\frac{\mu_B\cdot B}{\hbar}\cdot g_F m_F,
\end{split}
\end{equation}
where $\mu_B$ is the Bohr magneton, $B$ is the magnetic field amplitude, and $g_{J}$, $g_{F'}$ and $g_{F}$ are the Landé-factors for the Rydberg, the intermediate, and the ground states, respectively. For each $m_J$ transition path the EIT window should appear for a particular pump laser detuning, which depends on the energy shifts of all the three states~\cite{schlossberger2024zeeman}
\begin{equation}\label{eq:energy}
    \Delta_\mathrm{c}=\frac{\mu_B\cdot B}{\hbar}[g_{J}m_J-g_{F'}m_{F'}+\frac{f_c}{f_p}(g_{F'}m_{F'}-g_Fm_F)],
\end{equation}
where $f_p$ and $f_c$ are the frequencies of the probe and the pump fields, respectively. For the $\mathrm{^{87}Rb}$ $ \left|\mathrm{5S}_{1/2}, F=2, m_F=2\right> \rightarrow \left|\mathrm{5P}_{3/2}, F'=3, m_{F'}=3\right>$ $ \rightarrow \left|\mathrm{32D}_{5/2}, m_J=5/2\right>$ transition the pump field detuning, $\Delta_\mathrm{c}$, is equal to $3.6$~MHz/G. We were able to frequency tune the 480~nm laser by $-280$~MHz ($+280$~MHz) as shown on the Fig.~\ref{fig:lock}(a), when locked to the $m_J=-5/2$ ( $m_J=5/2$) peak, see Fig.~\ref{fig:lock}(b). The maximum magnetic field used was $\sim$73.1~G when a current of 10~A was driven through the coils.

\begin{figure}[ht!]
    \centering
    \includegraphics[width=8.5cm]{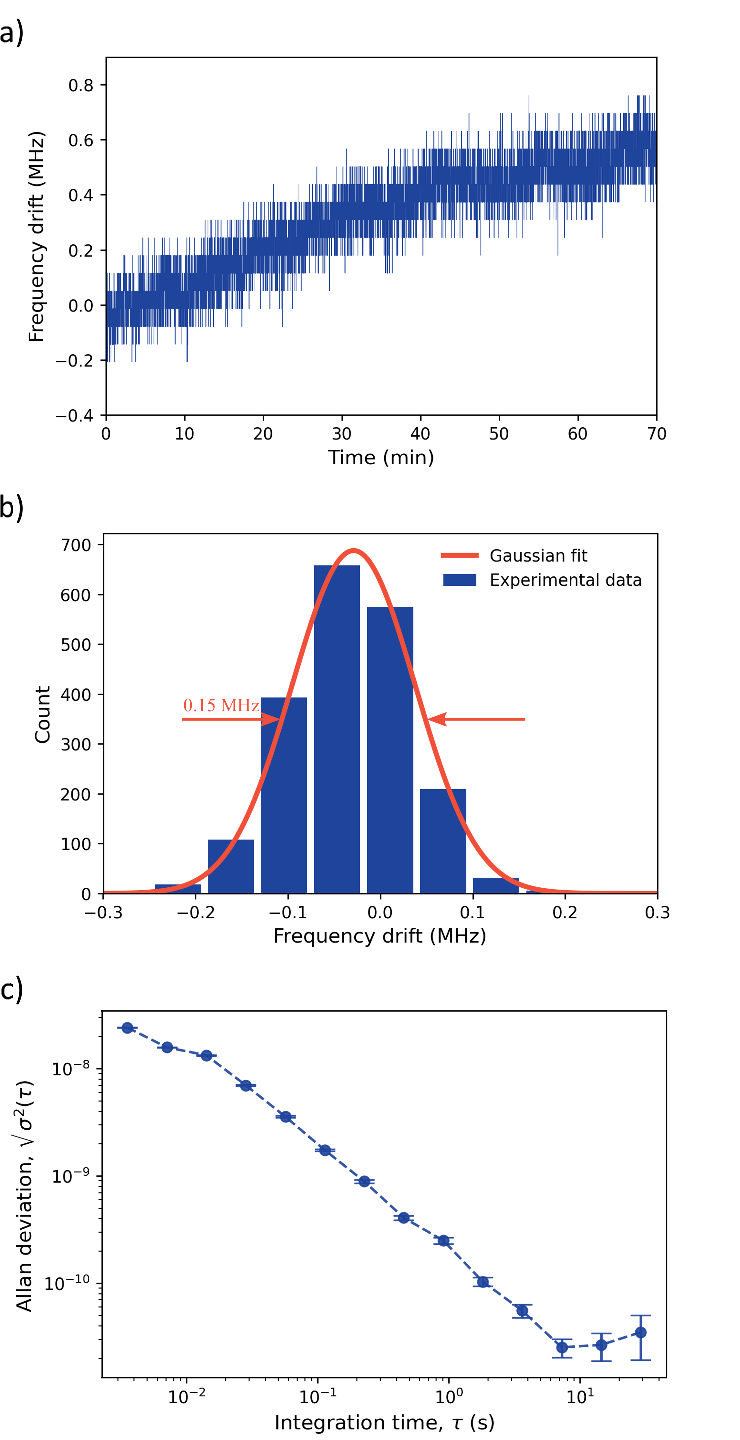}
    \caption{a) Frequency drift from the mean value of the laser locking over time when it is locked to the $\left|\mathrm{5P}_{3/2}, F'=3, m_{F'}=3\right>$ $ \rightarrow \left|\mathrm{32D}_{5/2}, m_J=-5/2\right>$ transition. b) Histogram of the frequency drift over 10 minutes recording time - blue, Gaussian fit of the histogram - red. The FWHM of 0.15~MHz is shown by the red arrows. c) Allan deviation, $\sqrt{\sigma^2(\tau)}$, versus the integration time, $\tau$. The minimum Allan deviation of $2.5\cdot10^{-11}$ is reached at the integration time $\tau\approx7$~s.}
    \label{fig:stability}
\end{figure}

The frequency stability of the 480~nm laser when locked to the $\left| 5\mathrm{P}_{3/2}, m_{F'}=3 \right> \rightarrow \left| 32\mathrm{D}_{5/2},m_J=-5/2 \right>$ transition was determined by recording its value for 70 minutes, see Fig.~\ref{fig:stability}(a). The frequency separation between the most distant mean frequency values gave an estimation of the long-term laser lock drift. Thence, we plotted a histogram of the frequency drift over a 10-minute recording, see Fig.~\ref{fig:stability}(b). We fitted the histogram with a Gaussian function and extracted the full-width-at-half maximum (FWHM) to estimate the short-term lock stability. The long-term frequency drift was $\approx0.5$~MHz, while the short-term lock stability was $\approx 0.15$~MHz, an improvement over that previously reported in the work of Rajasree~\textit{et al.}~\cite{rajasree20211}, where the long-term drift was $\sim 0.5$~MHz and the short-term stability was $\sim0.4$~MHz. We also characterized the frequency stability by means of Allan variance~\cite{riley2008handbook}, defined by $\sigma^2(\tau)=\frac{1}{2\tau^2}\sum^{N-2}_{n=1}(x_{n+2}-2x_{n+1}+x_n)^2$, where $x_n$ is the time series of the wavemeter frequency measurement spaced by the integration interval $\tau$. We observed a minimum Allan deviation, $\sqrt{\sigma^2(\tau)}$, on the order of $10^{-11}$ for an integration time of 7~s, see Fig.~\ref{fig:stability}(c). To address the stability change for different detunings of the 480 nm laser we measured the lock frequency using the wavemeter at two different magnetic field strengths and observed no change in the frequency drift. This could be explained by examining the amplitudes of the photodiode (PD) signal from the Zeeman-split EIT measurement. Once the $m_j=5/2$ and $m_j=-5/2$ peaks become distinguishable in the signal, their amplitudes show minimal variation as the magnetic field is increased, see Fig.~\ref{fig:lock}(b). In addition, we performed a heterodyne measurement by overlapping spatial modes of the target seed laser and a local oscillator via an optical fiber splitter. The local oscillator electric field was provided by a continuous wave (CW) Ti:Sapphire laser (MSquared), locked to a reference cavity yielding a spectral linewidth of 100 kHz. We introduced a frequency difference between the two lasers to observe a beat note signal. By performing a Fourier transform of the signal from the time to the frequency domain, we extracted a combined linewidth. Knowing the estimated linewidth of the local oscillator allowed us to calculate the linewidth of the target laser. When locking was performed in the absence of the external magnetic field, the linewidth was approximately 1.8 MHz. In contrast, applying the external magnetic field for the locking reduced the linewidth to about 0.8 MHz.

The EOM scan technique, reported in the work of Rajasree~\textit{et al.}~\cite{rajasree20211}, was limited in range by the Doppler width of the absorption spectrum of $\mathrm{^{87}Rb}$ where one could still observe the EIT signal. Therefore, the detuning of the probe laser could not exceed $\sim$450~MHz, or 0.8~GHz for the 480~nm laser. For the Zeeman level locking technique, the detuning of the probe field, $\Delta_{\mathrm{pr}}$, came from the energy shifts of the ground and intermediate state Zeeman levels and still had to be within the Doppler width. By inserting the Doppler limit of 450~MHz for the maximum probe frequency detuning into the following expression obtained from the eq.\ref{eq:levels}:
\begin{equation}
    \Delta_{\mathrm{pr}}=\frac{\mu_B \cdot B}{\hbar}\cdot(g_{F'}m_{F'}-g_Fm_F),
\end{equation}
we calculated the limit of the external magnetic field to be $\sim$324~G. By substituting this magnetic field into eq.~\ref{eq:energy} we estimated the limit for the pump field frequency detuning to be $\sim$1.17~GHz each side of the reference frequency. Hence the range of the scan could be extended to $\sim$2.3~GHz, one and half times larger than what was previously reported in the work of Rajasree~\textit{et al.}~\cite{rajasree20211}. However, using the EOM to tune the probe laser frequency to match the Zeeman shift when increasing the magnetic field, ensures that the  frequency remains within the Doppler-broadened absorption of the target Zeeman level. This, in principle, allows for arbitrary frequency scanning of the pump laser; however, in large magnetic fields (more than $\sim$500~G) the Paschen-Back effect should be considered~\cite{ma2017paschen}.

\section{Conclusion}
In conclusion, we have demonstrated a method to directly lock a 480~nm laser to the Zeeman sublevels of a Rydberg state, using the $\left|\mathrm{5P}_{3/2}, F=3\right> \rightarrow \left|\mathrm{32D}_{5/2}\right>$ transition as an example. This locking technique allows one to continuously tune the laser frequency over a range from -280~MHz to +280~MHz by controlling the strength of a transverse applied magnetic field. The scanning range could be increased further by increasing the maximum value of the applied magnetic field; however, in this case, care must be taken to avoid the quadratic Zeeman effect as the field gets larger~\cite{zhang2018interplay}. In the absence of an applied magnetic field, we have shown that the background magnetic fields introduce a polarization-dependent locking uncertainty due to the small Zeeman shift. The described technique is relatively low-cost and is useful for studies with Rydberg atoms where large continuous tuning of the laser frequency is required. The lock stability is comparable to or better than similar locking methods, with a short-term frequency stability of $0.15$~MHz and a long-term frequency drift of $0.5$~MHz recorded. We have also demonstrated that using the Rydberg Zeeman level reduces the lock linewidth by a factor of two. The limit of the 480~nm pump laser scanning range using the Rydberg Zeeman level, while the probe laser is locked to the 5S$_{1/2}\rightarrow 5$P$_{3/2}$ transition, is one and half times larger than that of a single-peak EIT locking scheme. The scanning range could be extended by heating the vapor cell to increase the Doppler width of the absorption spectrum for the 5S$_{1/2}\rightarrow 5$P$_{3/2}$ transition. In addition, the probe laser frequency can be matched with the Zeeman shift of a target laser via an EOM, in which case an arbitrary scanning range of the pump laser should be achievable.

\section*{Acknowledgements}
This work was supported by funding from the Okinawa Institute of Science and Technology Graduate University.  D.J.B. and S.N.C. acknowledge support from the Japan Society for the Promotion of Science (JSPS) Grant-in-Aid    No. 22K13986 (Early Career)  and No. 24K08289, respectively. 

\section*{Data Availability Statement}
The data is available from the corresponding authors upon reasonable request

\bibliography{aipsamp.bib}

\end{document}